\documentstyle[12pt,worldsci]{article}
\input epsf
\begin{document}
\vbox{\hfill\hbox{JLAB-THY 98--33}}
\title{Explicit Quark-Hadron Duality in 1+1 Dimensions\footnote{Talk
presented at the Third International Conference on Quark
Confinement and the Hadron Spectrum, Jefferson Lab, Newport News,
Virginia, June 7--12, 1998; to appear in Proceedings.}\footnote{Work
supported by the Department of Energy under contract No.\
DE-AC05-84ER40150.}}
\author{Richard F. Lebed\\
{\em Jefferson Lab, Newport News, Virginia 23606, USA}}
\maketitle
\setlength{\baselineskip}{2.6ex}

\vspace{0.7cm}
\begin{abstract}

	Explicit quark-hadron duality in the limit of heavy quark mass
is studied using the 't~Hooft model, where both partonic and hadronic
amplitudes may be computed exactly.  Results for weak decays of heavy
mesons are presented for both standard spectator decays, where the
duality limit is convincingly approached, and annihilation decays of
the valence quark-antiquark pair, where the approach to asymptotic
duality is much less precise.

\end{abstract}
\vspace{0.7cm}

\section{Introduction}

	One of the central issues in the study of confinement is
quark-hadron duality, the manner in which partial widths for exclusive
hadronic processes sum to the corresponding inclusive total at the
partonic level.  In this talk I present a brief summary of recent
advances\cite{GL1,GL2} in the study of this issue, obtained in
collaboration with Benjam\'{\i}n Grinstein.

	It is widely believed that the weak decay of a meson (``$\bar
B$'') containing a very heavy quark (``$b$'') should be well
approximated by the partonic decay of a free $b$.  From the physical
point of view, the heavy $b$ quark is indifferent to the dynamics of
the light spectator antiquark, and this reasoning is built into
nonrelativistic quark models, as well as heavy quark effective theory
and the ``practical version''\cite{SVZ} of the operator product
expansion (OPE), which expands in powers of $1/m_b$.  Exactly how this
``duality limit'' is achieved, or how closely it is satisfied, are
open questions.
\vspace{0.1cm}

\section{Method of Calculation}

	We address these issues in terms of a soluble toy theory
chosen to resemble real QCD as closely as possible.  The calculations
are performed in the context of the 't~Hooft model\cite{tH}, which is
the soluble theory of QCD in one space and one time dimension with an
infinite number $N_c$ of color charges.  Here ``soluble'' means that
each hadronic Green function (for example, meson wavefunctions and
transition amplitudes) may be calculated exactly (albeit numerically)
in terms of quark masses.  In particular, the $n$th meson
eigenfunction $\phi_n$ of mass eigenvalue $\mu_n$ for quarks of masses
$M$ and $m$ is found to satisfy
\begin{equation} \label{tHe}
\mu_n^2 \phi_n^{M\overline{m}} (x) = \left( \frac{M_R^2}{x} +
\frac{m_R^2}{1-x} \right) \phi_n^{M\overline{m}}(x) - \int^1_0 dy \,
\phi_n^{M\overline{m}} (y) \, \Pr \frac{1}{(y-x)^2},
\end{equation}
where masses are written in units of the gauge coupling constant $g
\sqrt{N_c/2\pi}$, since in 1+1 dimensions $g$ has dimensions of mass,
while in the large $N_c$ limit\cite{LargeNc} it scales as
$1/\sqrt{N_c}$.  The renormalized quark mass is then $m_R^2 = m^2 -
1$, while $x$, the sole remaining kinematic invariant, represents the
fraction of the meson momentum carried by the quark in light-cone
coordinates.

	Meson transition amplitudes, which are needed for decay
widths, may be written in terms of integral double and triple overlaps
of 't~Hooft wavefunctions.\cite{Ein} The nonleptonic decays
represented in these calculations are of the type $\bar B \to \pi_k
\pi_m$, where, {\it e.g.}, $\pi_k$ represents the $k$th radial
excitation of the pion.

	The partonic width, on the other hand, may be obtained through
a straightforward calculation leading to a closed-form expression in
terms of elliptic integrals with arguments that are functions of the
quark masses.  Results\cite{GL1} of the hadronic and partonic
calculations are compared in Fig.~1$a$.

	Physics in 1+1 dimensions possesses a number of peculiar
features, most notably the lack of spin (since spatial rotations are
absent), as well as singularities in two-body phase space at
threshold.  These singularities appear as spikes in Fig.~1$a$, but
remarkably do not affect the speed at which approximate duality is
achieved.

	That the exceptional agreement between hadronic and partonic
calculations is not accidental may be verified by studying the partial
width arising from each exclusive channel.  In particular, in Fig.1$b$
we exhibit the partial width for the decay of $\bar B$ into two
ground-state $\pi$ channels, which clearly does not saturate the
result of Fig.~1$a$.
\begin{figure} \label{fig1}
  \begin{centering}
	\def\epsfsize#1#2{0.38#2}
	\vbox{\hfil\epsfbox{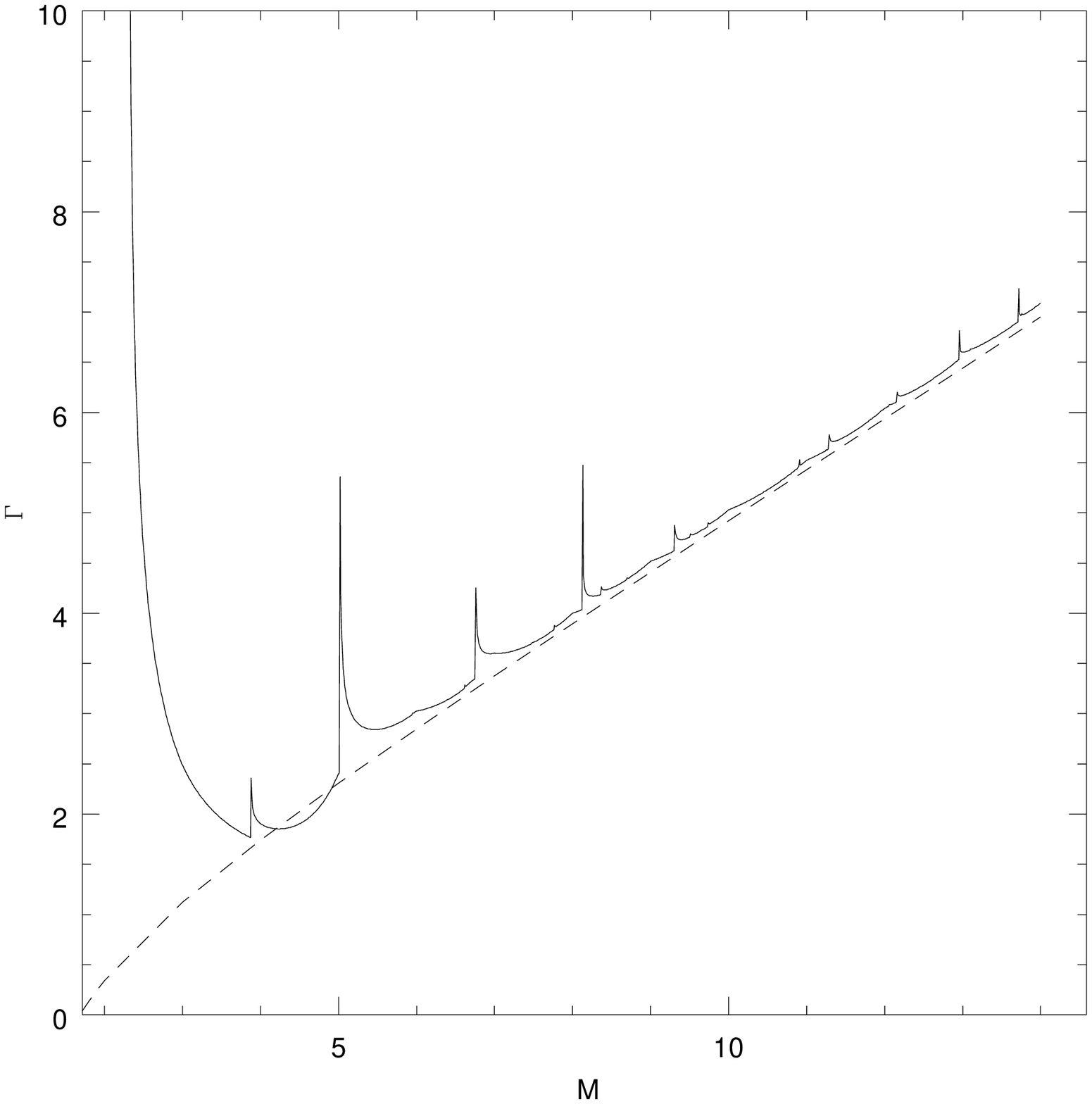}\hfill
	\epsfbox{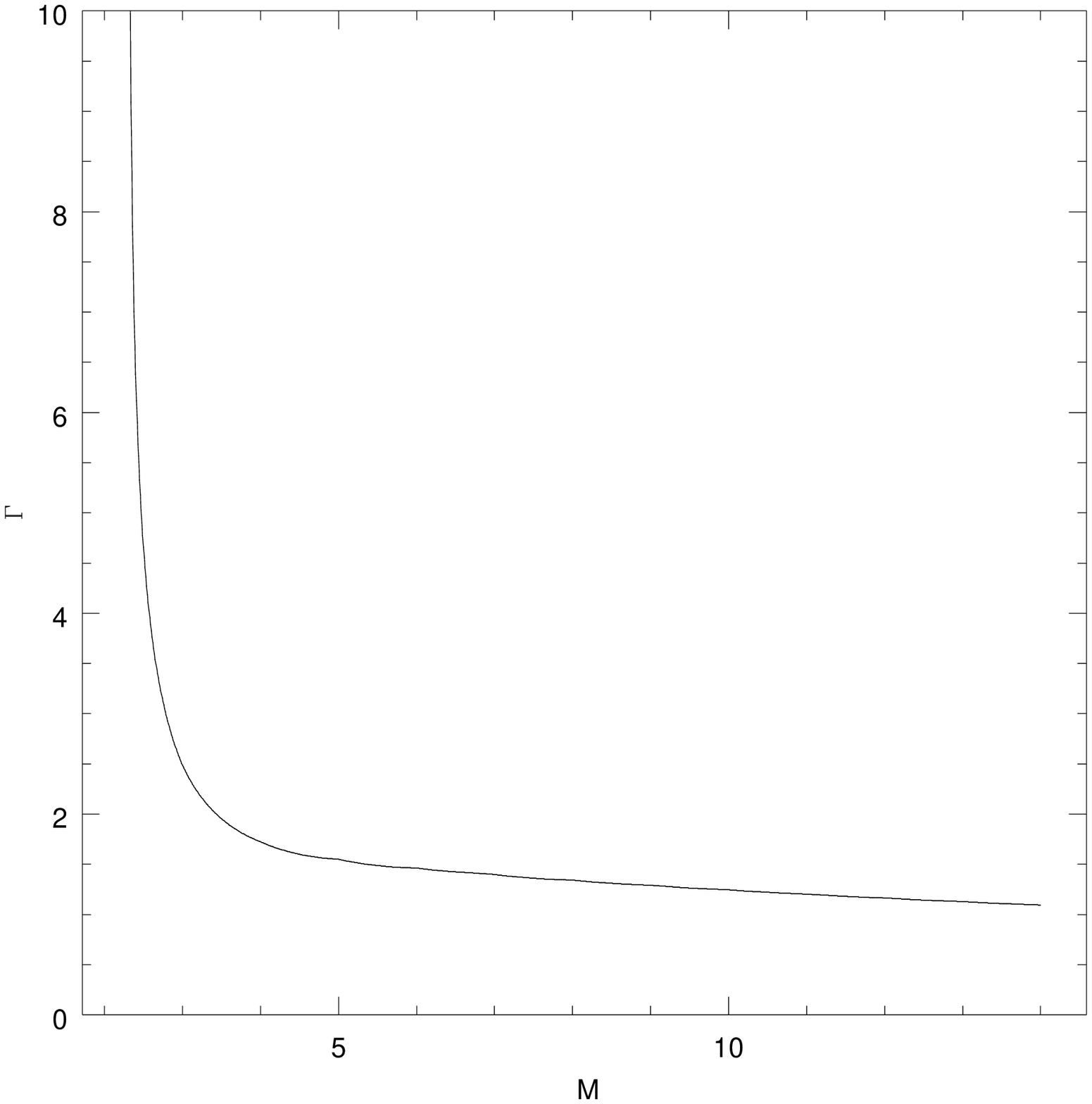}\hfil}

\caption{$(a)$ The full decay width for the sum of nonleptonic
exclusive modes in the decay of $\bar B$ as a function of heavy quark
mass $M$, with light quark mass $m = 0.56$.  The overall coefficient
in the width is suppressed for convenience.  The dashed line is the
tree-level parton result.  $(b)$ The partial hadronic width from $(a)$
in lowest exclusive channel.}

  \end{centering}
\end{figure}
%\vspace{0.3cm}
%

	The small discrepancy between the two curves at large $M$ has
received attention in the literature.  Numerically, it is well-fit by
a relative correction of order $1/M$, which should not be present in
expansions of the Ref.~\cite{SVZ} type.  Recent work\cite{Bigietal}
claims that such a correction is not present, partly on the basis of
rigorous analytic studies of 't~Hooft model solutions, but also partly
on estimates of scaling behavior in hadronic threshold regions.  The
issue, for the moment, appears to be unresolved; however, even if
Ref.\cite{Bigietal} proves to be absolutely correct, the presence of a
numerical effect {\em simulating\/} a forbidden correction may provide
valuable insight into processes in 3+1 dimensions that appear to
violate duality at $O(1/M)$, such as the lifetime difference between
$\Lambda_b$ and $\bar B$.
\vspace{0.1cm}

\section{Annihilation Decays}

	In the decays studied in the previous section, the light
spectator quark of the $\bar B$ is unaffected by the decay (except for
binding effects).  It is natural to consider also how well duality
holds for decays in which it weakly annihilates with the $b$
quark.\cite{GL2}

	In this case, one finds a new complication in that powers of
$N_c \to \infty$ no longer appear homogeneously in the amplitude.  The
reason for this behavior is clear: There are special values of quark
masses for which the $\bar B$ can annihilate resonantly into a single,
highly excited $\pi$ meson.  Since the presence of each additional
meson in large $N_c$ costs an extra factor of $\sqrt{N_c}$ in the
amplitude, one-meson decay widths should dominate the expected
two-meson widths by a factor of $N_c$.  However, this enhancement
appears only very near the special values of $b$ quark mass for which
such resonances are allowed, since large $N_c$ mesons have widths
$\Gamma_{\rm res} \propto 1/N_c$.

	In a physically meaningful picture, these resonances must be
``integrated out'' into the two-meson channel, where they appear
as intermediate mesons with Breit-Wigner widths in the form
\begin{equation}
\frac{i}{\mu_B^2 - \mu_{\rm res}^2 + i \mu_{\rm res} \Gamma_{\rm
res}}.
\end{equation}
The propagator, like meson masses, is $O(N_c^0)$ except in the
immediate vicinity of $\mu_B = \mu_{\rm res}$, where it is promoted to
$O(N_c^1)$.  It turns out that this is exactly the factor needed for
large $N_c$ counting consistency.  On the other hand, one finds that
different powers of $N_c$ are necessarily mixed up in obtaining a
physically reasonable picture of these decays, and a particular value
of $N_c$ must be chosen in order to exhibit numerical results.  Since
the 't~Hooft model is exact for $N_c = \infty$, $N_c$ is chosen as
large as numerically practical.

\begin{figure} \label{fig2}
  \begin{centering}
	\def\epsfsize#1#2{0.38#2}
	\vbox{\hfil\epsfbox{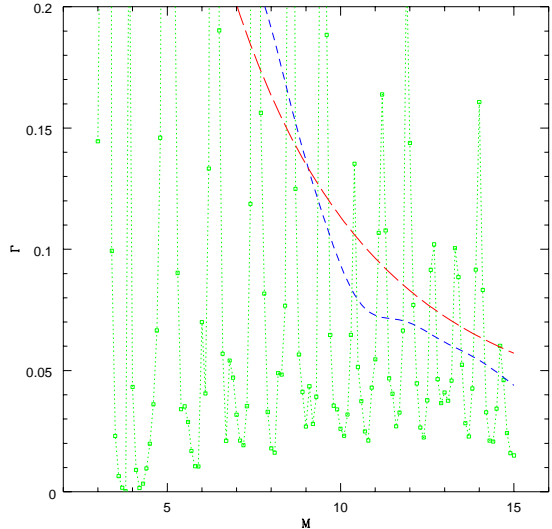}\hfil}

\caption{Total width of $\bar B$ meson through annihilation decays
(dotted line) as a function of heavy quark mass $M$, with $N_c =20$.
The width smeared by Gaussian averaging is in short dashes, while the
partonic result is in long dashes.}

  \end{centering}
\end{figure}
%\vspace{0.3cm}

	The results for $N_c=20$ are shown in Fig.~2, where the
prominent cusps in $\Gamma$ appear at values of $M$ for which
single-resonance decays can occur.  In order to compare to the
corresponding partonic rate, obviously some form of averaging, or
``smearing,'' process is required.  This is achieved here by weighting
the exact result by a series of Gaussian functions whose widths are
chosen as small as possible, but that still produce a smooth result.
Since larger $N_c$ means sharper spikes, while the range in $M$ over
which to smear is finite, numerically allowed choices of $N_c$ are
bounded above.  One then finds reasonable agreement between the two
curves, to about 20\%, but it is not nearly so startling as that
presented in Fig.~1$a$.  Also, the partonic and smeared hadronic
curves do not yet even appear to have achieved the same asymptotic
power law behavior.
\vspace{0.1cm}

\section{Conclusions}

	The 't~Hooft model provides a unique simplified environment in
which to study many of the currently intractable issues of the strong
interaction.  Even the issue of quark-hadron duality raises a number
of interesting subtleties, whose resolution may help in fathoming the
complexities of the 3+1 dimensional problem.


\vskip 0.9 cm
\thebibliography{References}

\bibitem{GL1}B. Grinstein and R. F. Lebed, Phys.\ Rev.\ D {\bf 57},
1366 (1998).

\bibitem{GL2}B. Grinstein and R. F. Lebed, Report No.\ UCSD/PTH 98-14
and JLAB-THY-98-18, hep-ph/9805404 (unpublished).

%\bibitem{HQET}N. Isgur and M. B. Wise, Phys.\ Lett.\ B {\bf 232}, 113
%(1989); B {\bf 237}, 527 (1990); E. Eichten and B. Hill, Phys.\ Lett.\
%B {\bf 234}, 511 (1990); M. B. Voloshin and M. A. Shifman, Yad.\ Fiz.\
%{\bf 47}, 801 (1988) [Sov.\ J. Nucl.\ Phys. {\bf 47}, 511 (1988)].

\bibitem{SVZ}M. A. Shifman, A. I. Voloshin, and V. I. Zakharov, Nucl.\
Phys.\ {\bf B147}, 385 and 448 (1979); V. Novikov, M. Shifman,
A. Vainshtein, and V. Zakharov, Nucl.\ Phys.\ {\bf B249}, 445 (1985).

\bibitem{tH}G. 't~Hooft, Nucl.\ Phys.\ {\bf B75}, 461 (1974).

\bibitem{LargeNc}G. 't Hooft, Nucl. Phys. {\bf B72}, 461 (1974).

\bibitem{Ein}M. B. Einhorn, Phys.\ Rev.\ D {\bf 14}, 3451 (1976);
B. Grinstein and P. F. Mende, Nucl.\ Phys.\ B {\bf 425}, 451 (1994).

\bibitem{Bigietal}I. Bigi, M. Shifman, N. Uraltsev and A. Vainshtein,
Minnesota U. Report No.\ TPI-MINN-97-29-T, hep-ph/9805241
(unpublished).
\end{document}